\documentstyle[11pt,amssymb,epsfig,psfig]{article}

\begin{document}

\title{Radiation from an oscillator uniformly moving along the axis
of a dielectric cylinder}
\author{A. A. Saharian\footnote{%
Email address: saharyan@server.physdep.r.am}  \\
\textit{Department of Physics, Yerevan State University, 375049
Yerevan, Armenia} \\
\textit{and Institute of Applied Problems in Physics, 375014
Yerevan, Armenia}\\
\textit{and} \\
A. S. Kotanjyan \\
\textit{Institute of Applied Problems in Physics, 375014 Yerevan,
Armenia}}

\date{\today}
\maketitle
\begin{abstract}
The radiation generated by a charged longitudinal oscillator
moving with a constant drift velocity along the axis of a
dielectric cylinder immersed in a homogeneous medium is
investigated. For an arbitrary oscillation law a formula is
derived for the spectral-angular distribution of this radiation.
Under the Cherenkov condition for the dielectric permittivity of
the external medium and oscillator drift velocity this formula
contains two summands. The first one corresponds to the radiation
with a continuous spectrum which propagates at the Cherenkov angle
of the external medium. The second one describes the radiation
which has a discrete spectrum for a given angle of propagation.
The corresponding frequencies are multiples of the Doppler-shifted
oscillation frequency. The results of numerical calculations for
the angular distribution of the radiated quanta are presented and
they are compared with the corresponding quantities for the
radiation in a homogeneous medium. It is shown that the presence
of the cylinder can increase essentially the radiation intensity.
\end{abstract}

\bigskip

PACS number(s): 41.60.-m, 41.60.Bq

\bigskip

\section{Introduction} \label{sec:oscint}

The operation of a number of devices assigned to production of
electromagnetic radiation is based on the radiation from
periodically moving charged particles, in particular, longitudinal
and transversal oscillators. Well-known examples are undulators,
wigglers and free-electron lasers (see, for instance, Refs.
\cite{Luch90,Brau90,Nikitin,Epp}). The extensive applications of
such devices motivate the importance of investigations for various
mechanisms of controlling the radiation parameters. From this
point of view, it is of interest to investigate the influence of
medium on the spectral and angular distributions of the radiation.
The presence of matter alters the radiation field and can give
rise to new types of radiation processes. This study is also
important with respect to some astrophysical problems.

Interfaces of media are widely used to control the radiation flow
emitted by various systems. Well-known examples of such a kind are
the Cherenkov radiation of a charge moving parallel to a plane
interface of two media or flying parallel to the axis of a
dielectric cylinder, transition radiation, Smith-Purcell
radiation. In Refs. \cite{Mkrt89} the radiation from a charged
particle flying over a surface acoustic wave generated on a plane
interface between two media is investigated. The corresponding
radiation from electron bunches and coherence effects are
considered in Ref. \cite{Saha01}. In a series of papers started in
Refs. \cite{Grigoryan1995,Grig95b} we have considered the most
simple geometries of boundaries, namely, the boundaries with
spherical and cylindrical symmetries. The radiation from a charge
rotating around a dielectric ball enclosed by a homogeneous medium
is investigated in Refs. \cite{Grig95b,Grigoryan1998}. It has been
shown that the interference between the synchrotron and Cherenkov
radiations leads to interesting effects: if for the material of
the ball and particle velocity the Cherenkov condition is
satisfied, there are strong narrow peaks in the radiation
intensity. At these peaks the radiated energy exceeds the
corresponding quantity in the case of a homogeneous medium by some
orders of magnitude. A similar problem for the case of the
cylindrical symmetry has been considered in Refs.
\cite{Grigoryan1995,Kot2000,Kot2001,Kota02,KotNIMB}. In Ref.
\cite{Grigoryan1995} a recurrent scheme is developed for
constructing the Green function of the electromagnetic field for a
medium consisting of an arbitrary number of coaxial cylindrical
layers. The investigation of the radiation from a charged particle
circulating around or inside a dielectric cylinder immersed in a
homogeneous medium, has shown that under the Cherenkov condition
for the material of the cylinder and the velocity of the particle,
there are narrow peaks in the angular distribution of the number
of quanta emitted into the exterior space. For some values of the
parameters the density of the number of quanta in these peaks
exceeds the corresponding quantity for the radiation in vacuum by
several orders.

In the present paper, on the basis of the Green function obtained
in Ref. \cite{Grigoryan1995}, the radiation by a longitudinal
oscillator with an arbitrary law of oscillations uniformly moving
along the axis of a dielectric cylinder is studied. The radiation
of charged oscillators in vacuum and inhomogeneous media was
considered in a number of papers (see, e.g., \cite{Epp},
\cite{KhachatryanB}-\cite{Bars87}, and references therein) in
connection with many possible applications (generation of
radiation, detection of high-energy particles, etc.). In
particular, the radiation from an oscillator with harmonic
oscillations in the laboratory frame in an infinite periodic
medium was studied in Refs. \cite{KhachatryanB,Kostanyan1971481}.
The spectral-angular distribution of the radiation intensity of an
oscillator moving over interface of two media was considered in
Ref. \cite{Kostanyan1977}. The radiation by a longitudinal
oscillator uniformly moving along the axis of a cylindrical
waveguide is investigated in Refs. \cite{Bars87,Saha03}. The
radiation properties in vacuum for the most general case of
harmonic oscillations in a four dimensional form are described in
Ref. \cite{Epp}. The special case of a longitudinal oscillator
with harmonic oscillations in the proper reference frame is
considered in Ref. \cite{Kuka75}.

This paper is organized as follows. In next section the
expressions for the vector-potential and electromagnetic fields
are derived for an arbitrary law of oscillation by using the Green
function. The angular-frequency distribution for the corresponding
radiation intensity is investigated in Sec. \ref{sec:oscraddistr}.
Two examples for the oscillatory motion corresponding to harmonic
oscillations in the laboratory and proper reference frames are
considered in Sec. \ref{sec:examples}. Section \ref{sec:Conc}
concludes the main results of the paper.

\bigskip

\section{Oscillator fields} \label{sec:oscfields}

Let a point charge $q$ moves along the axis of a dielectric
cylinder with permittivity $\varepsilon _0$ and radius $\rho _1$.
This system is immersed in a homogeneous medium with dielectric
permittivity $\varepsilon _1$. We consider the case when the
charged particle executes longitudinal oscillations with a drift
of the constant velocity $v_0$:
\begin{equation}
z=v_0t+f(t),\quad x=y=0,  \label{hetagic}
\end{equation}
where the Cartesian axis $z$ coincides with the cylinder axis, and
$f(t)$ is an arbitrary periodic function with the period $T=2\pi
/\omega _{0}$. In a properly chosen cylindrical coordinate system
($\rho ,\phi ,z$) the spatial part of the four-vector of the
current density created by this charge is written as
\begin{equation}
j_{l}=\frac{q}{\rho }v_{0}\delta (\rho )\delta (\phi )\delta (z-
v_{0}t-f(t))\delta _{lz,}\quad l=\rho ,\phi ,z.
\label{hosqxtutjun}
\end{equation}
The corresponding solution of the Maxwell equations for the
four-vector potential is expressed in terms of the Green function
$G_{il}({\mathbf{r}},t,{\mathbf{r}}^{\prime },t^{\prime })$ of the
electromagnetic field, which is a second-rank tensor:
\begin{equation}
A_{i}({\mathbf{r}},t)=-\frac{1}{2\pi ^{2}c}\int
G_{il}({\mathbf{r}},t,{\mathbf{r}}^{\prime },t^{\prime
})j_{l}({\mathbf{r}}^{\prime },t)d{\mathbf{r}}^{\prime }dt^{\prime
},\quad l=t,\rho ,\phi ,z.  \label{vektorpotencial}
\end{equation}
Taking into account the cylindrical symmetry of the problem under
consideration, it is convenient to present the Green function as a
Fourier expansion
\begin{equation}
G_{il}({\mathbf{r}},t,{\mathbf{r}}^{\prime }t^{\prime
})=\sum_{m=-\infty }^{\infty }\int_{-\infty }^{\infty
}dk_{z}d\omega G_{il}(m,k_{z},\omega ,\rho ,\rho ^{\prime })\exp
[im(\phi -\phi ^{\prime })+ik_{z}(z-z^{\prime })-i\omega
(t-t^{\prime })]. \label{GF_furieexp}
\end{equation}
For a dielectric cylinder with permittivity $\varepsilon _0$
immersed in a homogeneous medium with permittivity $\varepsilon
_1$, the Fourier transform of the Green function entering in Eq.
(\ref{GF_furieexp}) is given in Ref. \cite{Grigoryan1995}. When
$\rho ^{\prime }<\rho _1$, the function $ G_{il}(m,k_{z,}\omega
,\rho ,\rho ^{\prime })$ depends on $\rho ^{\prime }$ via the
Bessel function $J_m(\lambda _0\rho ^{\prime })$. For brevity,
here and below we use the notation
\begin{equation}
\lambda _j^2=\frac{\omega ^2}{c^2}\varepsilon _j-k_z^2,\quad j=0,1.
\label{lambdaj}
\end{equation}
Due to the presence of the function $\delta (\rho ^{\prime })$ in
expression (\ref{hosqxtutjun}) for the current density, only the
term with $m=0$ gives the nonzero contribution to the
vector-potential in Eq. (\ref{vektorpotencial}). As a result, the
field does not depend on the angle $\phi $. This is a simple
consequence of the azimuthal symmetry for the problem under
consideration. Thus, in this paper it is sufficient to have
expressions for $ G_{iz}(m=0,k_z,\omega ,\rho ,\rho ^{\prime
}=0)\equiv G_{iz}(k_z,\omega ,\rho )$ . In the Lorentz gauge these
expressions are immediately derived from the general formulas of
Ref. \cite{Grigoryan1995} and have the form
\begin{eqnarray}
G_{\rho z}(k_{z,}\omega ,\rho ) &=&-i\frac{(1-\varepsilon
_0/\varepsilon _1)k_zH_0(\lambda _1\rho _1)}{\rho _1\lambda
_0W_\varepsilon (J_0,H_0)W(J_0,H_0)}J_1(\lambda _0\rho
_{<})H_1(\lambda _1\rho _{>}),\, \rho _{_{>}^{<}}=
\begin{array}{c}
  \min \\
  \max
\end{array}
(\rho ,\rho _1), \nonumber  \label{GFcomponents} \\
G_{\phi z}(k_z,\omega ,\rho ) &=&0,  \\
G_{zz}(k_z,\omega ,\rho ) &=&\frac \pi {2iW(J_0,H_0)}\left\{
\begin{array}{ll}
W(J_{0},H_0)H_0(\lambda _0\rho )-W(H_0,H_0)J_0(\lambda _0\rho ),&
\rho <\rho _1, \\
2i H_0(\lambda _1\rho )/\pi \rho _1,& \rho >\rho _1,
\end{array}
\right.  \nonumber
\end{eqnarray}
where $H_m(x)=H^{(1)}_m(x)$ is the Hankel function of the first
kind. In these formulas the following notations are used:
\begin{eqnarray}
W(a,b) &=&a(\lambda _0\rho _1)\frac{\partial b(\lambda _1\rho _1)}{\partial
\rho _1}-b(\lambda _1\rho _1)\frac{\partial a(\lambda _0\rho _1)}{\partial
\rho _1} ,  \label{Wronskianner} \\
W_\varepsilon (J_0,H_0) &=&J_0(\lambda _0\rho _1)H_1(\lambda _1\rho _1)-%
\frac{\varepsilon _0\lambda _1}{\varepsilon _1\lambda _0}J_1(\lambda _0\rho
_1)H_0(\lambda _1\rho _1).  \nonumber
\end{eqnarray}
In the definition of $\lambda _1$ from Eq. (\ref{lambdaj}) one
should take into account that in the presence of the imaginary
part $\varepsilon _1^{\prime \prime }(\omega )$ for the dielectric
permittivity ($\varepsilon _1=\varepsilon _1^{\prime
}+i\varepsilon _1^{\prime \prime }$) the radiation field in the
exterior medium must damp exponentially for large $\rho $. This
leads to the following relations:
\begin{equation}
\lambda _1=\left\{
\begin{array}{c}
(\omega /c)\sqrt{\varepsilon _1-k_z^2 c^2/\omega ^2},\quad
\omega ^2\varepsilon _1/c^2>k_z^2, \\
i\sqrt{k_z^2-\omega ^2\varepsilon _1 /c^2},\quad \omega ^2
\varepsilon _1/c^2<k_z^2.
\end{array}
\right.  \label{lambda1}
\end{equation}
Note that for $\lambda _1^2>0$ the sign of $\lambda _1$ may be
also determined from the principle of radiation (different signs
of $\omega t$ and $\lambda _1\rho $ in expressions for the fields)
for large $\rho $.

Substituting Eq. (\ref{hosqxtutjun}) into formula
(\ref{vektorpotencial}), the following expressions are obtained
for the components of the vector-potential:
\begin{equation}
A_l(\rho ,z,t)=-\frac q{\pi c}\sum_{n=-\infty }^\infty
\int_{-\infty }^\infty dk_z\exp [i(zk_z-\omega
_n(k_z)t)]G_{lz}(k_z,\omega _n(k_z),\rho ) g_n(k_z) \frac{\omega
_n(k_z)}{k_z},  \label{Ai}
\end{equation}
where
\begin{equation}
\omega _n(k_z)=n\omega _0+k_zv_0,  \label{omegan}
\end{equation}
and $g_{n}(k_{z})$ is the Fourier transform of the function
$e^{ik_{z}f(t)}$:
\begin{equation} \label{gnkz}
e^{ik_{z}f(t)}=\sum_{n=-\infty }^{+\infty }g_{n}(k_{z})e^{in\omega
_{0}t},\quad
g_{n}(k_{z})=\frac{1}{T}\int_{-T/2}^{T/2}e^{ik_{z}f(t)-in\omega
_{0}t}dt.
\end{equation}
From the reality of the function $f(t)$ it follows that $%
g_{-n}(-k_{z})=g_{n}^{\ast }(k_{z})$. Fields inside and outside
the waveguide are derived by making use of the corresponding
expressions (\ref {GFcomponents}) for the components of the
reduced Green function. Below we will consider the fields in the
region $\rho
>\rho _{1}$. Taking into account expressions (\ref {GFcomponents}) and
formula (\ref{Ai}), for the vector-potential in this region one
obtains
\begin{equation}
A_{l}(\rho ,z,t)=\sum_{n=-\infty }^{+\infty }\int_{-\infty
}^{\infty }dk_{z}\exp [i(k_{z}z-\omega
_{n}(k_{z})t)]A_{nl}(k_{z},\rho ),\quad l=\rho ,z,\phi ,
\label{Arozetfi}
\end{equation}
where the coefficients $A_{nl}$ are determined by the relations
\begin{eqnarray}
A_{n\rho } &=&\frac{i q(1-\varepsilon _{0}/\varepsilon _{1})}{\pi
c\rho _{1}}\frac{\omega _{n}(k_{z})H_{0}(\lambda _{1}\rho
_{1})J_{1}(\lambda _{0}\rho _{1})}{\lambda _{0}W_{\varepsilon
}(J_{0},H_{0})W(J_{0},H_{0})}g_{n}(k_{z}) H_{1}(\lambda _{1}\rho ),
\nonumber \\
A_{nz} &=&-\frac{q}{\pi c\rho _{1}}\frac{\omega
_{n}(k_{z})g_{n}(k_{z})}{k_{z}W(J_{0},H_{0})}H_{0}(\lambda _{1}\rho ),  \label{Anrozetfi} \\
A_{n\phi } &=&0,  \nonumber
\end{eqnarray}
and in the definitions of $\lambda _j$ one has to substitute
$\omega =\omega _n(k_z)$. From the Lorentz gauge condition, by
using the symmetry of the problem under consideration, we obtain
for the scalar potential:
\begin{equation} \label{scpot}
\varphi _n(k_z,\rho )=\left[ \frac{1}{\rho }\frac{\partial (\rho
A_{n\rho })}{\partial \rho} +ik_z A_{nz}\right] \frac{c}{i\omega
_n(k_z)\varepsilon _1 }.
\end{equation}
As is seen from formula (\ref{Arozetfi}), analogous expressions
may also be written for the electric and magnetic fields. Having
expressions for the Fourier coefficients of the vector- and
scalar-potentials, one can derive the corresponding expressions
for these fields:
\begin{eqnarray}
E_{n\rho } &=&\frac{qg_n(k_z)}{\pi \rho _1\varepsilon _1W_\varepsilon
(J_0,H_0)}H_1(\lambda _1\rho ),\quad E_{nz}=\frac{iq\lambda _1g_n(k_z)}{\pi
\rho _1\varepsilon _1k_zW_\varepsilon (J_0,H_0)}H_0(\lambda _1\rho ),\quad
E_{n\phi }=0,  \label{electromagneticfield} \\
H_{n\phi } &=&\frac{q\omega _n(k_z)g_n(k_z)}{\pi c\rho
_1k_zW_\varepsilon (J_0,H_0)}H_1(\lambda _1\rho ),\quad H_{n\rho
}=H_{nz}=0.  \label{electromagneticfield1}
\end{eqnarray}
As follows from these formulas, ${\bf E}_{n}\cdot {\bf %
H}_{n}=0$, i.e., the Fourier components of the electric and
magnetic fields are perpendicular, and one can write
\begin{equation} \label{EHdir}
{\bf E}_{n}=[{\bf bH}_{n}], \quad {\bf b}=-\frac{c}{\omega
_{n}(k_{z})\varepsilon _{1}}(\lambda _{1},0,k_{z}).
\end{equation}
As functions of $k_z$ the Fourier coefficients for the fields
determined by relations (\ref{electromagneticfield}),
(\ref{electromagneticfield1}) have poles corresponding to the
zeros of the function $W_{\varepsilon }(J_0,H_0)$. It can be seen
that this function has zeros only for $\lambda _1^2<0<\lambda
_0^2$. As a necessary condition for this one has $\varepsilon
_1<\varepsilon _0$. Note that for the corresponding modes the
coefficients (\ref{electromagneticfield}),
(\ref{electromagneticfield1}) are proportional to the MacDonald
function $K_m(|\lambda _1|\rho )$, $m=0,1$, and they are
exponentially damped with the distance from the cylinder axis.
These modes are precisely the eigenmodes of the dielectric
cylinder and propagate inside the dielectric cylinder. Below, in
the consideration of the intensity for the radiation to the
exterior medium, we will disregard the contribution of the poles
corresponding to these modes.

\section{Spectral-angular distribution of the radiation intensity
} \label{sec:oscraddistr}

The average energy flux per unit time through the cylindrical
surface of radius $\rho $ coaxial with the dielectric cylinder is
given by the Poynting vector ${\bf S}$ as
\begin{equation}
I=\frac{2\pi }T\int_0^Tdt\int_{-\infty }^\infty ({\bf Sn}_\rho
)\rho dz,\quad {\bf S=}\frac c{4\pi }[{\bf EH}],\quad T=\frac{2\pi
}{\omega _0}, \label{Poyntingvector}
\end{equation}
where ${\bf n_\rho}$ is the unit vector of the cylindrical
surface. By making use of formulas (\ref{Arozetfi}),
(\ref{electromagneticfield}), and (\ref{electromagneticfield1}),
we obtain
\begin{equation}
I=-\frac{i q^2\rho }{\pi \rho _1^2}\sum_{n=-\infty }^{+\infty
}\int_{-\infty
}^\infty dk_z\frac{\lambda _1\omega _n(k_z)\left| g_n(k_z)\right| ^2}{%
\varepsilon _1k_z^2\left| W_\varepsilon (J_0,H_0)\right|
^2}H_0(\lambda _1\rho )H_1^{*}(\lambda _1^*\rho ).
\label{intensivut}
\end{equation}
Using the fact that the replacement $n\rightarrow -n$,
$k_z\rightarrow -k_z$ leads to $\omega _n(k_z)\rightarrow -\omega
_n(k_z)$, $\lambda _1\rightarrow -\lambda _{1}^*$ and $\left|
g_{-n}(-k_z)\right| ^2=\left| g_n(k_z)\right| ^2$, expression
(\ref{intensivut}) can be written in the explicitly real form
\begin{equation}
I=\frac{2q^2\rho }{\pi \rho _1^2}\sum_{n=0}^\infty {}^{\prime }\int_{-\infty
}^\infty dk_z\frac{\omega _n(k_z)\left| g_n(k_z)\right| ^2}{%
\varepsilon _1k_z^2\left| W_\varepsilon (J_0,H_0)\right|
^2}{\mathrm{Im}} [\lambda _1 H_0(\lambda _1\rho )H_1^{*}(\lambda
_1^{*}\rho )], \label{intensreal}
\end{equation}
where the prime over the sum means that the term with $n=0$ must
be taken with the weight $1/2$. Now by using Eq.(\ref{lambda1})
and the Wronskian for the Bessel functions, it can be easily seen
that
\begin{equation}\label{Im2}
  {\mathrm{Im}} [\lambda _1 H_0(\lambda _1\rho )H_1^{*}(\lambda
_1^{*}\rho )=\left\{
\begin{array}{ll}
2/\pi \rho,& \lambda_1^{2}>0,
\\ 0,& \lambda_1^{2}<0.
\end{array}
\right.
\end{equation}
Hence, for the energy flux one obtains
\begin{equation}\label{fluxnew}
  I=\frac{4q^2}{\pi^2 \rho _1^2}\sum_{n=0}^\infty {}^{\prime }
\int_{\lambda_1^{2}>0}dk_z\frac{\omega _n(k_z)\left| g_n(k_z)\right| ^2}{%
\varepsilon _1k_z^2\left| W_\varepsilon (J_0,H_0)\right| ^2}.
\end{equation}
As we could expect in the absence of absorption this flux does not
depend on the distance $\rho$ from the cylinder axis.

We first discuss the case $\omega _0=0$, which corresponds to the
uniform motion of the charge along the axis of a dielectric
cylinder. Then, according to Eqs. (\ref{lambdaj}) and
(\ref{omegan}), $\omega _n(k_z)$ and $\lambda _j$ do not depend on
$n$, and the series in Eq. (\ref{intensreal}) is easily summarized
by the standard formula for the Fourier transformation:
\begin{equation}
\sum_{n=-\infty }^{+\infty }\left| g_n(k_z)\right| ^2=\frac
1T\int_{-T/2}^{T/2}\left| e^{ik_zf(t)}\right| ^2dt=1.
\end{equation}
For $\omega_0=0$ one has $\lambda_j^{2}=k_z^{2}(\beta_j^{2}-1)$,
where $\beta_j=v_0\sqrt{\varepsilon_j}/c$, $j=0,1$, and flux
(\ref{fluxnew}) is nonzero if the Cherenkov condition for the
external medium is fulfilled, $\beta _1>1$.  Passing to
integration over $\omega =k_z$v$_0$, we obtain
\begin{equation}
I=\frac{4q^2v_0}{\pi ^2\rho _1^2}\int_{\beta _1^2>1}
d\omega\frac{1 }{\varepsilon _1\omega \left| W_\varepsilon
(J_0,H_0)\right| ^{2}}, \quad \omega _0=0 .  \label{Cerenkovglan}
\end{equation}
It is easy to show that this expression coincides with the formula
presented, for example, in Ref. \cite{Bolotovsky} for the
radiation of a charge moving parallel to the axis of a cylindrical
channel in a dielectric. Introducing the angle $\vartheta $ of the
wave vector with the cylinder axis, from the relation $\omega
=k_z$v$_0$ we obtain $\cos \vartheta =\beta _1^{-1}$, i.e., the
radiation described by Eq. (\ref{Cerenkovglan}) propagates under
the Cherenkov angle of the external medium. When $\varepsilon
_0=\varepsilon _1$, from Eq. (\ref{Cerenkovglan}) we obtain the
well-known formula for the intensity of the Cherenkov radiation in
a homogeneous medium.

Now consider the  general case, when $\omega _0\neq 0$. First we
analyze the term with $n=0$ in formula (\ref{fluxnew}).
Analogously to deriving of Eq. (\ref{Cerenkovglan}), it is easy to
show that the corresponding contribution to the radiation
intensity is determined by the formula
\begin{equation}
I_{n=0}=\frac{4q^2v_0}{\pi ^2\rho _1^2}\int_{\beta
_1^2>1}d\omega\frac{ \left| g_0(\omega /v_0)\right| ^{2}
}{\varepsilon _1\omega \left| W_\varepsilon (J_0,H_0)\right| ^2}.
\label{Intn0}
\end{equation}
This radiation propagates under the Cherenkov angle of the
external medium and its intensity differs from the intensity of
the uniformly moving charge, given by Eq. (\ref{Cerenkovglan}), by
the presence of the frequency-dependent additional weight factor
$\left| g_0(\omega /v_0)\right| ^2$.

Now we proceed to the terms with $n\neq 0$ in formula
(\ref{fluxnew}). From the condition $\lambda_1^{2}>0$ we have the
following quadratic inequality with respect to $k_z$:
\begin{equation}
k_z^2(1-\beta _1^{-2})+2k_z\frac{n\omega _0}{v_0}+
\frac{n^2\omega _0^2}{v_0^2%
}>0.  \label{qarakusihav}
\end{equation}
Let the Cherenkov condition be not fulfilled initially for the
drift velocity of the charge: $\beta _1<1$. In this case we obtain
from this inequality:
\begin{equation}
k_z\in \left( -\frac{n\omega _0\sqrt{\varepsilon _1}}{c(1+\beta _1)},\frac{%
n\omega _0\sqrt{\varepsilon _1}}{c(1-\beta _1)}\right) .  \label{kazettirujt}
\end{equation}
It is convenient to introduce a new variable $\vartheta $
according to
\begin{equation}
k_z=\frac{n\omega _0}c\frac{\sqrt{\varepsilon _1}\cos \vartheta
}{1-\beta _1\cos \vartheta },  \label{kazet}
\end{equation}
where from relation (\ref{kazettirujt}) it follows that $\vartheta
\in (0,\pi )$. Then from expression (\ref{omegan}) we have
\begin{equation}
\omega _n(k_z)=\frac{n\omega _0}{1-\beta _1\cos \vartheta },\
n=1,2... \label{omegatet}
\end{equation}
Note that in accordance with Eqs. (\ref{kazet}) and
(\ref{omegatet}) the quantities $k_z$ and $\omega _n(k_z)$ are
connected by the relation $k_z=\omega _n(k_z)\sqrt{\varepsilon
_1}\cos \vartheta /c$.

Now consider the case $\beta _1>1$, when the solution of
inequality (\ref{qarakusihav}) has the following form:
\begin{equation}
k_z\in \left( -\infty ,-\frac{n\omega _0\sqrt{\varepsilon _1}}{c(\beta _1-1)}%
\right) \cup \left( -\frac{n\omega _0\sqrt{\varepsilon _1}}{c(\beta _1+1)}%
,\infty \right) .  \label{kazettirujt1}
\end{equation}
Introducing again the variable $\vartheta $ according to
expression (\ref{kazet}), we ensure that $\vartheta \in
(0,\vartheta _{0})\cup (\vartheta _{0},\pi )$, where $\vartheta
_0=\arccos (\beta _1^{-1})$ is the corresponding Cherenkov angle
for the drift velocity. Relations between the variables  $k_z$,
$\omega _n(k_z)$, and $ \vartheta $ now are the same as for $\beta
_1<1$. At large distances from the charge trajectory the
dependence of elementary waves on the space time coordinates has
the form $\exp [\omega _n(k_z)\sqrt{ \varepsilon _1}(\rho \sin
\vartheta +z\cos \vartheta -ct/\sqrt{\varepsilon _1 })/c]$, which
describes the wave with the frequency
\begin{equation}
\omega _n=\left| \omega _n(k_z)\right| =\frac{n\omega _0}{|1-\beta _1\cos
\vartheta |},  \label{omegantet}
\end{equation}
propagating at the angle $\vartheta $ to the  $z$-axis. Formula
(\ref{omegan}) describes the normal Doppler effect in the cases
$\beta _1<1$ and $\beta _1>1$, $\vartheta >\vartheta _0$ and
anomalous Doppler effect in the case $\beta _1>1$, $\vartheta
<\vartheta _0$. By making use of the formulas given above, the
expressions for the $\lambda _j$ can be written as
\begin{equation}
\lambda _0=\frac{n\omega _0}{c}\frac{ \sqrt{\varepsilon
_0-\varepsilon _1\cos ^2\vartheta }}
{1-\beta _1\cos \vartheta }, \quad \lambda _1=%
\frac{n\omega _0}{c}\frac{\sqrt{\varepsilon _1}\sin \vartheta
}{1-\beta _1\cos \vartheta }.   \label{lambosc}
\end{equation}
Taking into account these remarks and passing from integration
over $k_z$ to integration over $\vartheta $, we obtain the
following expression for the radiation intensity:
\begin{equation}
I_{n\neq 0}=\sum_{n=1}^\infty \int \frac{dI_n}{d\Omega }d\Omega
,\quad \frac{ dI_n}{d\Omega }=\frac{2q^2c}{\pi ^3\rho
_1^2\varepsilon _1^{3/2}}\frac{\left| g_n(nu)\right| ^2}{\left|
W_\varepsilon (J_0,H_0)\right| ^2\left| 1-\beta _1\cos \vartheta
\right| \cos ^2\vartheta },  \label{dIndomega}
\end{equation}
where $d\Omega =\sin \vartheta d\vartheta d\phi $ is the element
of solid angle and the notation
\begin{equation}
u=\frac{\omega _0}{c}\frac{\sqrt{\varepsilon _1}\cos \vartheta }{
1-\beta _1\cos \vartheta  }  \label{u}
\end{equation}
is introduced. To obtain the radiation intensity at the angle
$\vartheta =\pi /2$, note that for $n\neq 0$ and $k_z\to 0$ from
formula (\ref{gnkz}) one has
\begin{equation}\label{gnkzto0}
g_n(k_z)\approx i k_z f_n,\quad f_n=\frac{1}{T}\int_{-T/2}^{T/2}
f(t)e^{-i n \omega _0 t}dt \ ,
\end{equation}
with $f_n$ being the Fourier transform of the function $f(t)$. Now
from Eqs. (\ref{dIndomega}) and (\ref{u}) one finds
\begin{equation}\label{dIndomtetpi2}
\frac{ dI_n}{d\Omega }=\frac{2q^2\omega _0^2}{\pi ^3 c \rho
_1^2\sqrt{\varepsilon _1}} \frac{n^2 |f_n|^2}{|W_{\varepsilon
}(J_0,H_0)|^2} , \quad \lambda _j=\frac{n\omega
_0}{c}\sqrt{\varepsilon _j}, \quad j=0,1\,\quad
\vartheta=\frac{\pi}{2}\ .
\end{equation}
By taking into account Eqs. (\ref{kazet}), (\ref{omegatet}), and
(\ref{lambosc}), from formula (\ref{EHdir}) one obtains the
following relation between the electric and magnetic fields for
the radiation at $n$th harmonic:
\begin{equation} \label{polariz}
{\mathbf{E}}_n=\frac{1}{\sqrt{\varepsilon _1}}\left[
{\mathbf{H}}_n {\mathbf{n}} \right] ,
\end{equation}
where ${\mathbf{n}}=(\sin \vartheta ,0, \cos \vartheta )$ is the
unit vector in the propagation direction. Hence, the radiation is
linearly polarized and the polarization plane goes through the
vectors ${\mathbf{n}}$ and ${\mathbf{v}}_0$.

The total intensity of the radiation for the longitudinal
oscillator moving with a constant drift velocity along the axis of
a dielectric cylinder can be written as the sum
\begin{equation} \label{totintOm}
I=I_{n=0}+I_{n\neq 0}\, ,
\end{equation}
where the first term on the right is given by formula
(\ref{Intn0}) and describes the radiation with a continuous
spectrum propagating at the Cherenkov angle of the external
medium, if the condition $\beta _1>1$ is fulfilled (for $\beta
_1<1$ this term is absent). The second term describes the
radiation, which for a given angle $\vartheta $ has a discrete
spectrum determined by formula (\ref{omegantet}). With allowance
for the dispersion of the dielectric permittivity, this term does
not contribute to the radiation at the Cherenkov angle. This is
connected with the fact that for a given $n>0$ the frequency
defined by Eq. (\ref{omegantet}), tends to infinity as $\vartheta
$ approaches $\vartheta _0$ and hence beginning from a certain
frequency the Cherenkov condition ceases to be fulfilled. The
angles for which the dispersion should be taken into account, are
determined implicitly from the condition $\omega _n\geq \omega _d$
by using formula (\ref{omegantet}) and frequency dependence of the
permittivity $\varepsilon _1=\varepsilon _1(\omega _n)$, where
$\omega _d$ is the characteristic frequency of the dispersion.

Substituting $\varepsilon _1=\varepsilon _0$, from the formulas
given above we obtain the corresponding expressions for an
oscillator moving in a homogeneous medium. In particular, by
taking into account the formula
\begin{equation}
W_{\varepsilon }(J_0,H_0)=-\frac{2ic(1-\beta _1\cos \vartheta
)}{\pi \rho _1 n\omega _0 \sqrt{\varepsilon _1}\sin \vartheta },
\quad \varepsilon _0=\varepsilon _1, \label{Wepshom}
\end{equation}
from Eq. (\ref{dIndomega}) for the corresponding radiation
intensity one obtains
\begin{equation}
I_{n\neq 0}^{(0)}=\sum_{n=1}^\infty \int \frac{dI_n}{d\Omega
}d\Omega ,\quad \frac{dI_n^{(0)}}{d\Omega }=\frac{q^2n^2\omega
_0^2}{2\pi c\sqrt{\varepsilon _1}}\frac{\tan ^2\vartheta \left|
g_n(nu)\right| ^2}{\left| 1-\beta _1\cos \vartheta \right| ^3}.
\label{dIndomegahamaser}
\end{equation}

Let us consider the radiation intensity given by Eq.
(\ref{dIndomega}) for large values of the harmonic $n$. The
behaviour of the function $g_n(n u)$ for large $n$ can be
estimated by using the stationary phase method. To do this we note
that this function has the form
\begin{equation}\label{gnnew2}
  g_n(n u)=\frac{1}{T}\int_{-T/2}^{T/2}e^{inS(t)}dt,\quad S(t)=u
  f(t)-\omega _0t \, .
\end{equation}
The asymptotic behaviour of the integral for large values $n$,
$n\to \infty $, is essentially different in dependence wether the
function $S(t)$ has a stationary point, being the solution of the
equation $S'(t)=0$. By using expression (\ref{u}) for $u$, this
equation can be written in terms of the particle velocity $v(t)$
as
\begin{equation}\label{statpoint}
  \frac{v(t)}{c}\sqrt{\varepsilon _1}\cos \vartheta =1, \quad
  v(t)=v_0+f'(t).
\end{equation}
It follows from here that, if for the particle velocity $v(t)$ the
Cherenkov condition is not fulfilled, the subintegrand in Eq.
(\ref{gnnew2}) has no stationary points and for $f(t)\in C^{\infty
}$ the integral is exponentially suppressed for large values of
the harmonic number $n$. When the Cherenkov condition is fulfilled
the main term of the asymptotic expansion of integral in Eq.
(\ref{gnnew2}) is determined by the contribution of the stationary
point and in accordance with the standard formula in the
stationary phase method, one has $g_n(n u)\propto 1/\sqrt{n}$,
$n\to \infty $. In this case in the absence of the dispersion the
radiation intensity linearly increases with increasing $n$ and in
addition to the radiation from an oscillating charge one has also
the Cherenkov radiation. To identify the angular regions where the
Cherenkov radiation propagates, let us denote by $v_{+}$ and
$v_{-}$ maximal and minimal values of the function $v(t)$ and
\begin{equation}\label{betplmin2}
  \beta _{\pm }=v_{\pm }\sqrt{\varepsilon _1}/c.
\end{equation}
The angular regions where the Cherenkov radiation propagates are
as follows:
\begin{eqnarray}\label{cases}
  && \vartheta < \vartheta _{+}, \, \, \vartheta _{+}\equiv
  \arccos \frac{1}{\beta _{+}} \quad
  {\mathrm{for}} \quad  -1<\beta _{-}<1<\beta _{+} \nonumber \\
  && \vartheta _{-}<\vartheta <\vartheta _{+}, \, \, \vartheta _{-}\equiv
  \arccos \frac{1}{\beta _{-}} \quad {\mathrm{for}} \quad \beta _{-}>1 \\
  && \vartheta >\vartheta _{-}, \quad {\mathrm{for}} \quad \beta
  _{-}<-1,\quad \beta _{+}<1 \nonumber \\
  && \vartheta <\vartheta _{+}\cup \vartheta >\vartheta _{-}\quad
{\mathrm{for}} \quad \beta _{+}>1,\quad \beta _{-}<-1 . \nonumber
\end{eqnarray}
In the first two cases the Cherenkov radiation propagates in the
inward direction to the $z$-axis. In the last two cases the
Cherenkov radiation in the region $\vartheta >\vartheta _{-}$
propagates in the backward direction. As we have mentioned, the
radiation intensity in regions (\ref{cases}) linearly increases
with increasing $n$. However, note that large values $n$
correspond to large frequencies for which the dispersion becomes
important. Hence, for the radiation in regions (\ref{cases}) the
value of the harmonic $n$ for which the radiation intensity has a
maximum is determined by the dispersion through the conditions
$\beta _{+}>1$ or $|\beta _{-}>1|$. Here the situation is similar
to the case of the ordinary Cherenkov radiation.

By taking into account formula (\ref{dIndomegahamaser}) for the
radiation in a homogeneous medium with permittivity $\varepsilon
_1$, expression (\ref {dIndomega}) for the radiation intensity at
$n$th harmonic in the presence of a dielectric cylinder may be
written in the following form
\begin{equation}
\frac{dI_n}{d\Omega }=\frac{dI_n^{(0)}}{d\Omega }F_n(\vartheta ),
\label{dIn}
\end{equation}
where the inhomogeneity factor is given by the formula
\begin{equation}
F_n(\vartheta )=\frac{4c^2\left| W_\varepsilon (J_0,H_0)\right|
^{-2}}{\pi ^2\rho _1^2\varepsilon _1\omega _n^2\sin ^2\vartheta }.
\label{Fntet}
\end{equation}
Consider the dependence of this factor on the cylinder radius in
the limit $\left| \lambda _j\right| \rho _1\gg 1$, when the
wavelength for the radiation is much less than the cylinder
radius. Using the asymptotic expressions for the cylindrical
functions for large values of the argument and taking into account
formulas (\ref{lambosc}), for the inhomogeneity factor we get
\begin{equation}
F_n(\vartheta )=\frac{2\sqrt{\varepsilon _{0/}\varepsilon _1-\cos
^2\vartheta }}{\sin \vartheta }\left[ 1+\frac{\varepsilon _0^2\sin
^2\vartheta }{\varepsilon _1^2(\varepsilon _0/\varepsilon _1-\cos
^2\vartheta )}+\left( 1-\frac{\varepsilon _0^2\sin ^2\vartheta
}{\varepsilon _1^2(\varepsilon _0/\varepsilon _1-\cos ^2\vartheta
)}\right) \sin (2\left| \lambda _0\right| \rho _1)\right] ^{-1}
\label{Fntet1}
\end{equation}
for $\cos ^2\vartheta <\varepsilon _0/\varepsilon _1$ ($\lambda
_0^2>0$) and
\begin{equation}
F_n(\vartheta )=\frac{4\sqrt{\cos ^2\vartheta -\varepsilon
_{0/}\varepsilon
_1}}{\sin \vartheta }e^{-2\left| \lambda _0\right| \rho _1}\left[ 1+\frac{%
\varepsilon _0^2\sin ^2\vartheta }{\varepsilon _1^2(\cos
^2\vartheta -\varepsilon _0/\varepsilon _1)}\right] ^{-1}
\label{Fntet2}
\end{equation}
for $\cos ^2\vartheta >\varepsilon _0/\varepsilon _1$ ($\lambda
_0^2<0$). As we see, in the second case the intensity
exponentially decreases with increasing $\rho _1$. This is caused
by the fact that for $\varepsilon _0/\varepsilon _1<1$ the angle
$\arccos \sqrt{\varepsilon _{0/}\varepsilon _1}$ corresponds to
the angle of total internal reflection, and in the limit of the
geometric optics the beams incident from inside on the cylinder
surface cannot propagate at the angles $\cos ^2\vartheta
<\varepsilon _0/\varepsilon _1$ in the external medium.

\begin{figure}[tbph]
\begin{center}
\begin{tabular}{cc}
\epsfig{figure=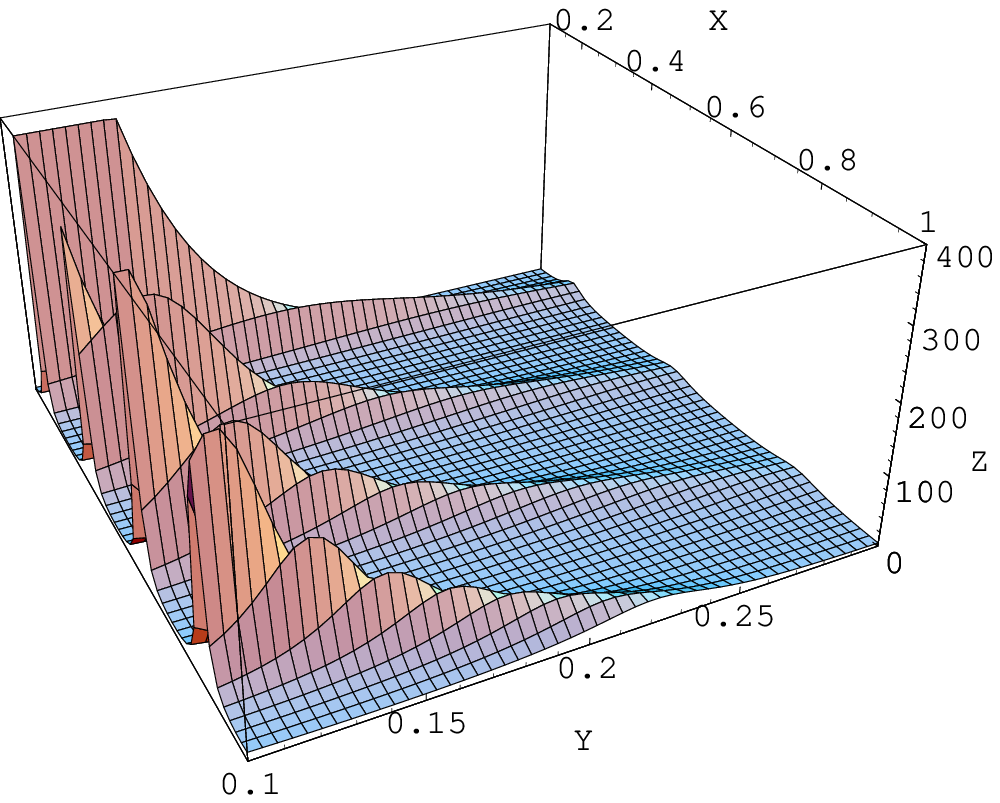,width=6.5cm,height=6cm}& \quad
\epsfig{figure=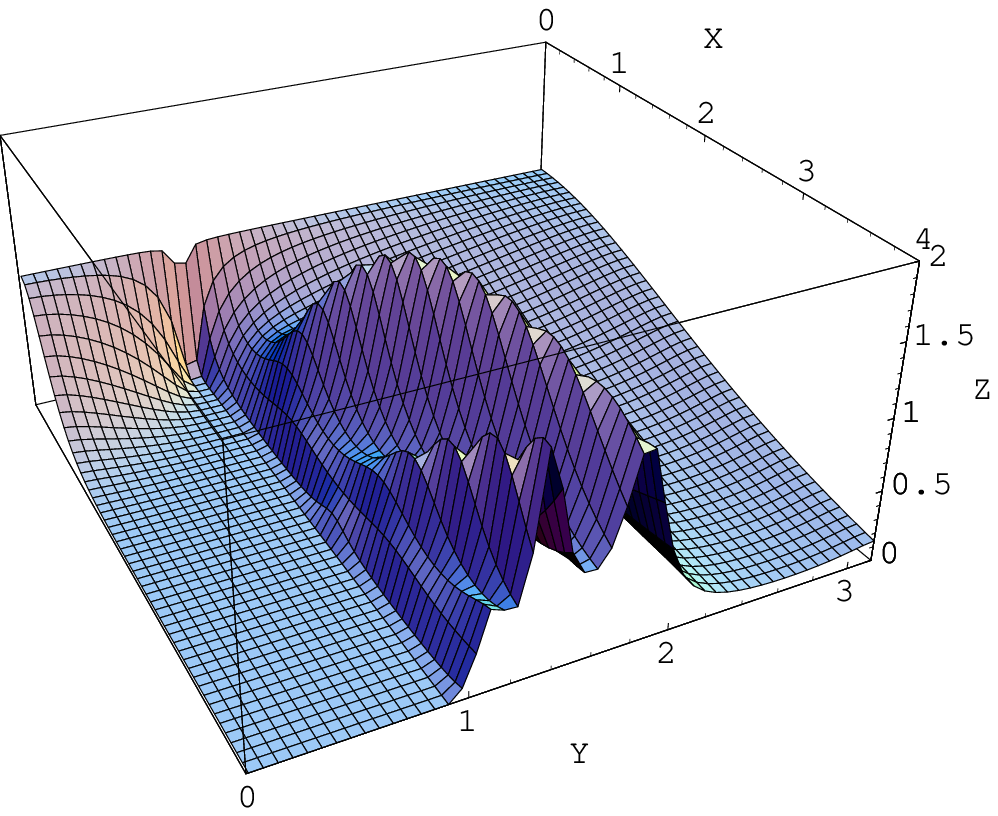,width=6.5cm,height=6cm}
\end{tabular}
\end{center}
\caption{The inhomogeneity factor $Z=F_{n}$ as a function on
$X=n\omega _{0}\rho _1/c$ and $Y=\vartheta $ for $v_0=0.9c$. Left
figure corresponds to the values $\varepsilon _0=3$, $\varepsilon
_1=1 $ and right figure corresponds to $\varepsilon _0=1$,
$\varepsilon _1=3 $. } \label{fig1}
\end{figure}

In the opposite limit, when the wavelength for the radiation is
much larger than the cylinder radius $|\lambda _j |\rho _1 \ll 1$,
using the asymptotic expressions for cylindric functions, it is
easy to show that from Eq. (\ref{dIndomega}) one can immediately
derive the radiation intensity in a homogeneous medium with the
permittivity $\varepsilon _1$ (formula (\ref{dIndomegahamaser})),
i.e., $F_n(\vartheta )\to 1$. For fixed values of $\varepsilon_j$
and $v_0$ the inhomogeneity factor $F_n$ is a function on
$n\omega_0\rho_1/c$ and $\vartheta $:
\begin{equation}\label{FN2}
  F_n=F\left(\frac{n\omega_0\rho_1}{c},\vartheta\right) .
\end{equation}
In Fig. \ref{fig1} we have plotted this function for the charge
drift velocity $v_0=0.9c$ in two different cases:
$\varepsilon_0=3$, $\varepsilon_1=1$ (left figure) and
$\varepsilon_0=1$, $\varepsilon_1=3$ (right figure). As we see, in
the first case the presence of cylinder can essentially increase
the radiation intensity. In the left graph it is clearly seen the
exponential suppression of the inhomogeneity factor with
increasing $n\omega_0\rho_1/c$ for angles $\vartheta <\arccos
\sqrt{\varepsilon _0/\varepsilon _1}\approx 0.96$.

\section{Examples of the oscillatory motion} \label{sec:examples}

In this section we will consider two examples of the oscillation
law, corresponding to the harmonic oscillations in the laboratory
and proper reference frames.

\subsection{Harmonic oscillation in the laboratory frame}
\label{subsec:ex1}

First of all let us consider the case when the charged particle
oscillates as an harmonic oscillator in the laboratory frame with
the amplitude $z_0$:
\begin{equation}\label{fsinomt}
f(t)=z_0 \sin (\omega _0 t).
\end{equation}
From formula (\ref{gnkz}) it can be easily seen that for this type
of motion one has
\begin{equation}\label{gnkzsint}
g_n(k_z)=J_n(k_z z_0).
\end{equation}
Hence, in this case the radiation intensity is given by formulas
(\ref{Intn0}) and (\ref{dIndomega}) with the function $g_n(k_z)$
given by Eq. (\ref{gnkzsint}). In particular, for $\vartheta =\pi
/2$ the radiation intensity is nonzero only at the fundamental
harmonic $n=1$. Due to the well-known properties of the Bessel
function $J_n(nz)$, the behavior of the radiation intensity as a
function of $n$ depends essentially on the sign of the quantity
$1-(uz_0)^2$. By using formula (\ref{u}), this quantity can be
presented in the form
\begin{equation}\label{w}
  1-w^2=\frac{1-\beta _{-}\cos \vartheta }{(1-\beta _1 \cos \vartheta )
  ^2}(1-\beta _{+}\cos \vartheta ) ,\quad w=uz_0,\
\end{equation}
where we use notation (\ref{betplmin2}) with
\begin{equation}\label{betapm1}
  v _{\pm }=v_0\pm \omega _0z_0 .
\end{equation}
If the Cherenkov condition for the maximum velocity of the charge
is not fulfilled, $\beta _{+}<1$, one has $w<1$. By using Debye's
asymptotic expansion for the function $J_n(nw)$, it can be seen
that for large values of the harmonic $n$ the radiation intensity
behaves as
\begin{equation}\label{Ilargen}
  I_n\sim \frac{n}{x} \left( \sqrt{\frac{1-x}{1+x}}e^x \right)
  ^{2n}, \quad x=\sqrt{1-w^2}.
\end{equation}
It follows from here that for $x\simeq 1$ the spectral
distribution has maximum at
\begin{equation}\label{nmax}
  n\simeq n_{{\mathrm{max}}}=(1-w^2)^{-3/2},
\end{equation}
and is exponentially suppressed for larger harmonics.

For $\beta _{+}>1$ regions for values of the angle $\vartheta $
exist for which $w>1$. In these regions in addition to the
radiation from an oscillating charge, one has also the Cherenkov
radiation. The angular regions where the Cherenkov radiation
propagates are determined by formulas (\ref{cases}). As follows
from the formula for the radiation intensity in combination with
Eq. (\ref{gnkzsint}), for a given harmonic $n$ the corresponding
intensity vanishes at angles determined from the equation $|u
z_0|=j_{n,l}/n,$ $ l=1,2,...,$ where $j_{n,l}$ are the positive
zeros of the Bessel function: $J_n(j_{n,l})=0$. If the Cherenkov
condition in the external medium for the maximum velocity of the
charge $v_0+\omega_0z_0$ is not fulfilled, then $|u z_0| <1$, and
as follows from the relation $j_{n,l}>n$ for the zeros of the
Bessel function, this equation has no solutions. If the Cherenkov
condition is fulfilled, then the radiation intensity at a given
harmonic $n$ vanishes at the angles $\vartheta $ determined from
the formula
\begin{equation}
\cos \vartheta _{n,l}=\frac c{\sqrt{\varepsilon _1}(v_0\pm n\omega
_0z_0/j_{n,l})}.  \label{costetnl}
\end{equation}
These angles do not depend on the dielectric permittivity of the
cylinder and, hence, are the same for the radiation in the
presence of the cylinder and in a homogeneous medium.

Consider now formula (\ref{dIndomegahamaser}) for the radiation
intensity in a homogeneous medium with the function $g_n(k_z)$
from (\ref{gnkzsint}). First we analyze the case when the
Cherenkov condition is not fulfilled for the maximal velocity of
the charge, $\beta _{+}<1$, when $\left| u z_0 \right| <1$. If we
neglect the dispersion, then in Eq. (\ref{dIndomegahamaser}) the
series over $n$ is summarized by means of the standard formula
\cite{Prudnikov2}
\begin{equation}
\sum_{n=1}^\infty n^2J_n^2(nw)=\frac{w^2(w^2+4)}{16(1-w^2)^{7/2}},
\quad w<1 \, . \label{Spravgum}
\end{equation}
Then from Eq. (\ref{dIndomegahamaser}) for the angular
distribution of the radiation intensity we obtain
\begin{equation}
\frac{dI^{(0)}}{d\vartheta }=\frac{q^2\omega _0^2}{16c\sqrt{\varepsilon_1}}%
\frac{\tan ^2\vartheta \sin \vartheta }{\left| 1-\beta _1\cos
\vartheta \right| ^3}\frac{w^2(w^2+4)}{(1-w^2)^{7/2}}.
\label{dIn0}
\end{equation}
For $\varepsilon _1=1$ this formula can be obtained from the
general formula for the radiation of a four-dimensional harmonic
oscillator presented, for instance, in Refs. \cite{Nikitin,Epp}.
For $\beta _{+}>1$, a range of values of $\vartheta $ exists,
where $\left| u z_0 \right|>1$ (see Eqs. (\ref{cases})), and in
the absence of dispersion the series over $n$ in Eq.
(\ref{dIndomegahamaser}) diverges. Thus, in this case, in order to
obtain a finite result for the radiation intensity, the
consideration of the dispersion is necessary. In the limit of low
velocities of the oscillatory motion ($\omega _0 z_0 \ll c$) the
radiation intensity for a given harmonic at angles different from
the Cherenkov one tends to zero as $(\omega _0 z_0 / c)^{2n}$ for
both homogeneous and inhomogeneous cases.

\begin{figure}[tbph]
\begin{center}
\begin{tabular}{cc}
\epsfig{figure=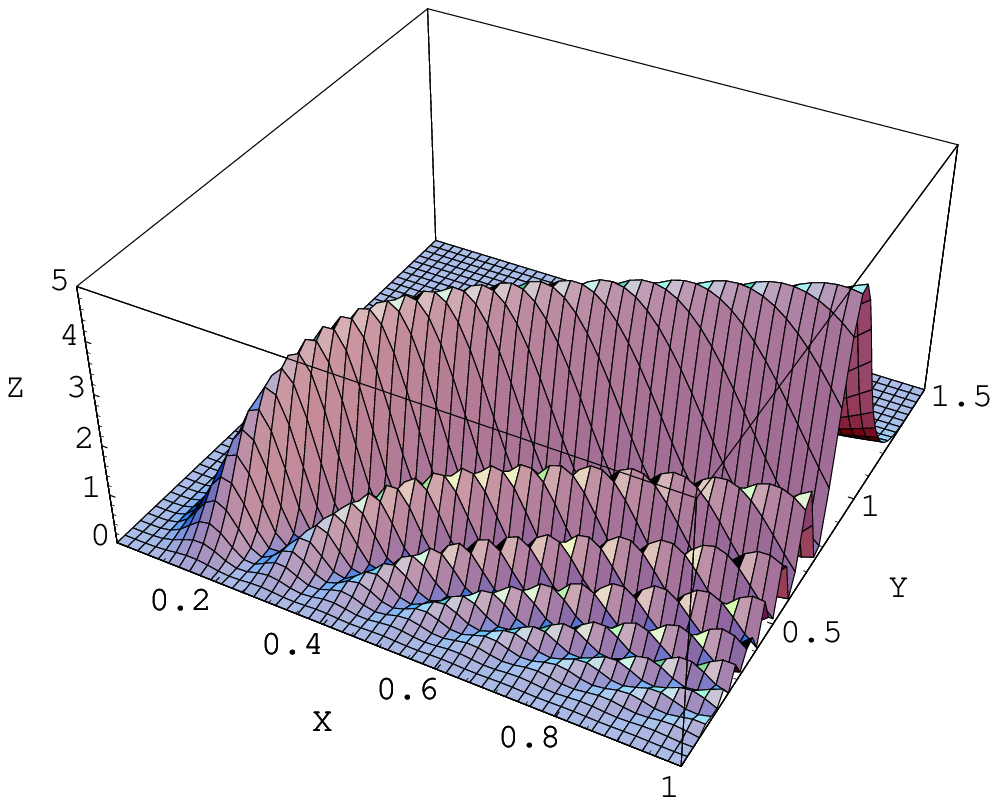,width=6.5cm,height=6cm}& \quad
\epsfig{figure=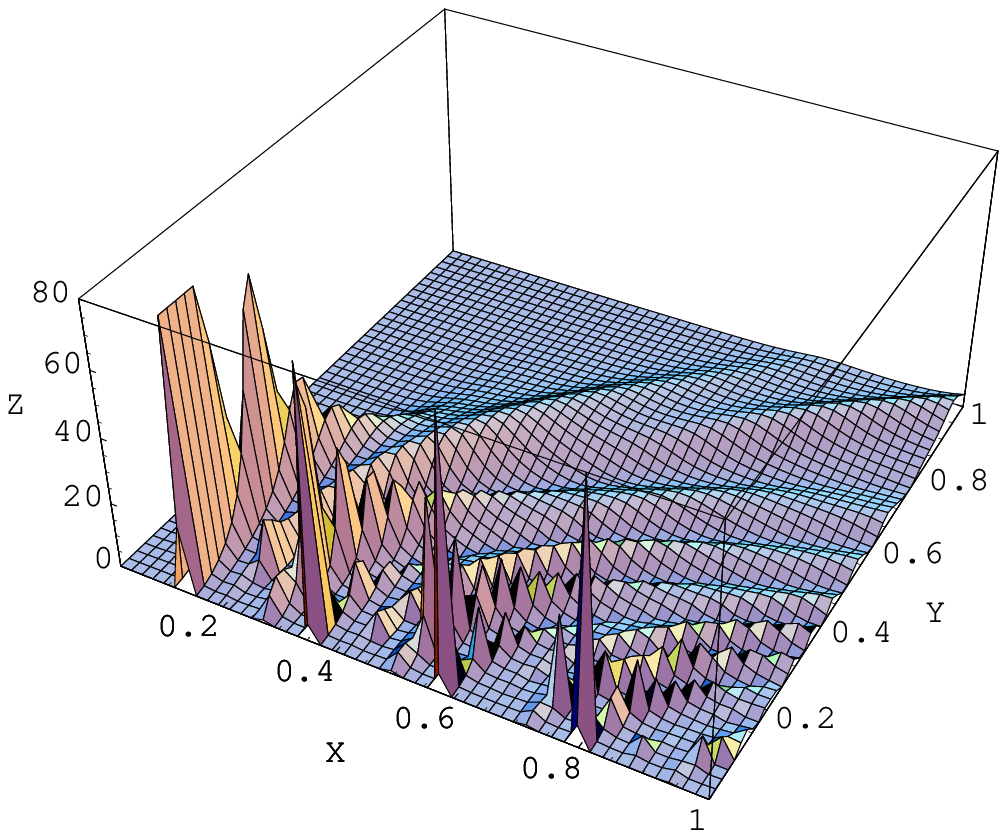,width=6.5cm,height=6cm}
\end{tabular}
\end{center}
\caption{Number of emitted quanta per solid angle, $Z=(\hbar
c\sqrt{\varepsilon_1}/q^2)(2\pi /\omega _0)dN_n/d\Omega $, as a
function on $X=\omega _0 z_0\sqrt{\varepsilon _1}/c $ and
$Y=\vartheta $ for $v_0\sqrt{\varepsilon _1}/c=0.9$, $n=3$ in a
homogeneous medium, $\varepsilon _0=\varepsilon _1$ (left figure)
and in presence of a cylinder with $\varepsilon _0/\varepsilon
_1=3$, $\rho _1=z_0$ (right figure).} \label{fig3mass}
\end{figure}
\begin{figure}[tbph]
\begin{center}
\epsfig{figure=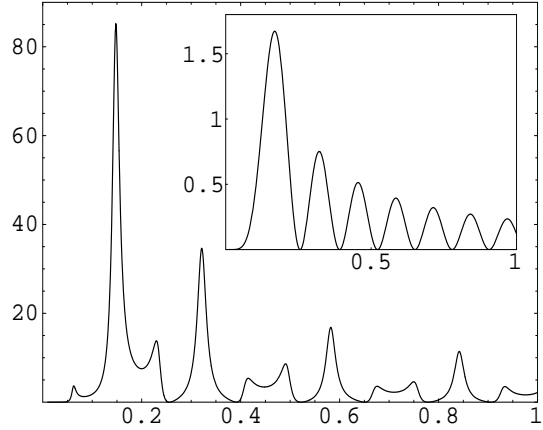,width=9cm,height=6cm}
\end{center}
\caption{The same quantity as in Fig. \ref{fig3mass} as a function
on $\omega _0 z_0\sqrt{\varepsilon _1}/c $ for $\vartheta =0.2 $,
$v_0\sqrt{\varepsilon _1}/c=0.9$, $n=3$, $\varepsilon
_0/\varepsilon _1=3$, $\rho _1=z_0$. The graph in the right-up
corner corresponds to the same quantity for the oscillator
radiation in a homogeneous medium ($\varepsilon _0=\varepsilon
_1$). }\label{fig3}
\end{figure}

We have performed numerical calculation for the angular
distribution of the density of number of quanta
\begin{equation}\label{numerquant1}
  \frac{dN_{n}}{d\Omega}=\frac{1}{\hbar\omega_{n}}\frac{dI_{n}}{d\Omega}
\end{equation}
for several values of the harmonic number $n$. Figures
\ref{fig3mass} and \ref{fig3} show the results of these
calculations for the oscillator radiation in a homogeneous medium
and in presence of a cylinder with dielectric permittivity
$\varepsilon_{0}/\varepsilon _1=3$. As is seen from these figures
the presence of the cylinder can essentially increase the
radiation intensity.

\subsection{Harmonic oscillation in the proper reference
frame} \label{subsec:ex2}

In this subsection we consider the radiation from a charge that
executes harmonic oscillations in the reference frame which moves
at the constant velocity $v_0$. In the comoving coordinate system,
the coordinates of the particle are given by
\begin{equation}\label{properosc}
z'=a' \sin (\omega '_0 t'), \quad x'=y'=0.
\end{equation}
The radiation generated by this kind of oscillator in the vacuum
(the case $\varepsilon _0=\varepsilon _1=1$) has been studied in
Ref. \cite{Kuka75} (see also Ref. \cite{Epp}). In the laboratory
reference frame the motion of the oscillator is described by
formula (\ref{hetagic}), where the function $f(t)$ is defined in
the parametric form as
\begin{equation}\label{fsintprime}
f(t)=a'\sqrt{1-\beta ^2}\sin (\omega '_0 t'), \quad
t=\frac{1}{\sqrt{1-\beta ^2}}\left( t' +\frac{a'v_0}{c^2}\sin
(\omega '_0 t') \right)\, \quad \beta =\frac{v_0}{c}\, .
\end{equation}
This function is periodic with the period $T=2\pi /\omega _0$,
where $\omega _0 $ is related to the proper frequency $\omega '_0$
by the standard formula $\omega _0=\omega '_0 \sqrt{1-\beta ^2}$.
For the function defined by Eq. (\ref{fsintprime}) one has
$f(-t)=-f(t)$, and the function $g_n(k_z)$ is real. Substituting
into Eq. (\ref{gnkz}) expressions (\ref{fsintprime}) and
performing to the new integration variable $x=\omega '_0 t'$, one
finds
\begin{equation}\label{gnkzprime}
g_n(k_z)=\frac{1}{\pi }\int_{0}^{\pi }(1+A\cos x)\cos (n B \sin x
-nx ) dx \ ,
\end{equation}
where
\begin{equation}\label{AB}
A=\frac{a' \omega '_0 v_0}{c^2},\quad B=\frac{k_z
a'}{n}\sqrt{1-\beta ^2} - A.
\end{equation}
Note that using formula (\ref{kazet}), the expression for the
coefficient $B$ can be presented in the form
\begin{equation}\label{B}
B=\frac{\omega '_0 a'}{c \sqrt{\varepsilon _1}} \frac{\varepsilon
_1 \cos \vartheta -\beta _1}{1-\beta _1\cos \vartheta }.
\end{equation}
The integral in Eq. (\ref{gnkzprime}) can be evaluated by using
the standard formula from Ref. \cite{Prudnikov2}:
\begin{equation}\label{gnkzprime2}
g_n(k_z)=J_n(nB) \frac{(1-\beta ^2) \varepsilon _1 \cos \vartheta
}{\varepsilon _1 \cos \vartheta -\beta _1}.
\end{equation}
Substituting this into general formula (\ref{dIndomega}), for the
angular distribution of the radiation intensity at $n$th harmonic
one finds
\begin{equation}\label{dIndomegac2}
\frac{dI_n}{d\Omega }=\frac{2q^2 c \sqrt{\varepsilon _1}}{\pi
^3\rho _1^2}\frac{(1-\beta ^2)^2 J_n^2(nB)}{|W_\varepsilon
(J_0,H_0)|^2 |1-\beta _1\cos \vartheta | (\varepsilon _1 \cos
\vartheta -\beta _1)^2} \ .
\end{equation}
By taking into account formula (\ref{Wepshom}), for the radiation
intensity of an oscillator moving in a homogeneous medium with
dielectric permittivity $\varepsilon _1$,   one obtains
\begin{equation} \label{Inthom2}
\frac{d I^{(0)}_{n}}{d\Omega }=\frac{q^2 \varepsilon
_1^{3/2}}{2\pi c} \frac{n^2\omega _0^2(1-\beta ^2)^2\sin
^2\vartheta J_n^2(nB)}{|1-\beta _1\cos \vartheta |^3(\varepsilon
_1\cos \vartheta -\beta _1 )^2}.
\end{equation}
For $\varepsilon _1=1$ this expression coincides with the formula
derived in Ref. \cite{Kuka75}.

As in the previous example, the behavior of the radiation
intensity (\ref{dIndomegac2}) is essentially different for the
cases $B<1$ and $B>1$. Note that from the expression for $B$ it
follows that
\begin{equation}\label{B1}
1-B^2=\frac{1-(v_0\omega '_0a'/c^2)^2}{(1-\beta _1\cos \vartheta
)^2} \left( 1-\beta _{-}\cos \vartheta \right) \left( 1-\beta
_{+}\cos \vartheta \right) \ ,
\end{equation}
where $\beta _{\pm }$ are defined by Eq. (\ref{betplmin2}) with
\begin{equation}\label{vplus}
v_{\pm }=\frac{v_0 \pm \omega '_0a'}{1\pm v_0\omega '_0a'/c^2} \,
.
\end{equation}
If for the maximal velocity of the charge, $v_{+}$, the Cherenkov
condition is not satisfied, $\beta _{+}<1$, from (\ref{B1}) one
has $B<1$ and the radiation intensity behaves as in Eq.
(\ref{Ilargen}) with the replacement $w\to B$. In this case in the
corresponding formula for the radiation in a homogeneous medium,
given by Eq. (\ref{Inthom2}), the series over $n$ can be
summarized with the help of formula (\ref{Spravgum}) assuming that
the dispersion is absent:
\begin{equation}\label{dIndomegahamc2}
\frac{dI^{(0)}}{d\Omega }=\sum_{n=1}^{\infty
}\frac{dI_n^{(0)}}{d\Omega } =\frac{q^2\omega _0^2
\sqrt{\varepsilon _1} }{32\pi c} \left(\frac{\omega
'_{0}a'}{c}\right) ^2 \frac{(1-\beta ^2)^2\sin ^2 \vartheta
}{|1-\beta _1\cos \vartheta |^5} \frac{(4+B^2)}{(1-B^2)^{7/2}} \,
,
\end{equation}
where $1-B^2$ is given by Eq. (\ref{B1}). For $\beta _{+}>1$
regions for $\vartheta $ exist with $B>1$. In these regions in
addition to the radiation from an oscillating charge one has also
the Cherenkov radiation. These regions are determined by relations
(\ref{cases}).
\begin{figure}[tbph]
\begin{center}
\begin{tabular}{cc}
\epsfig{figure=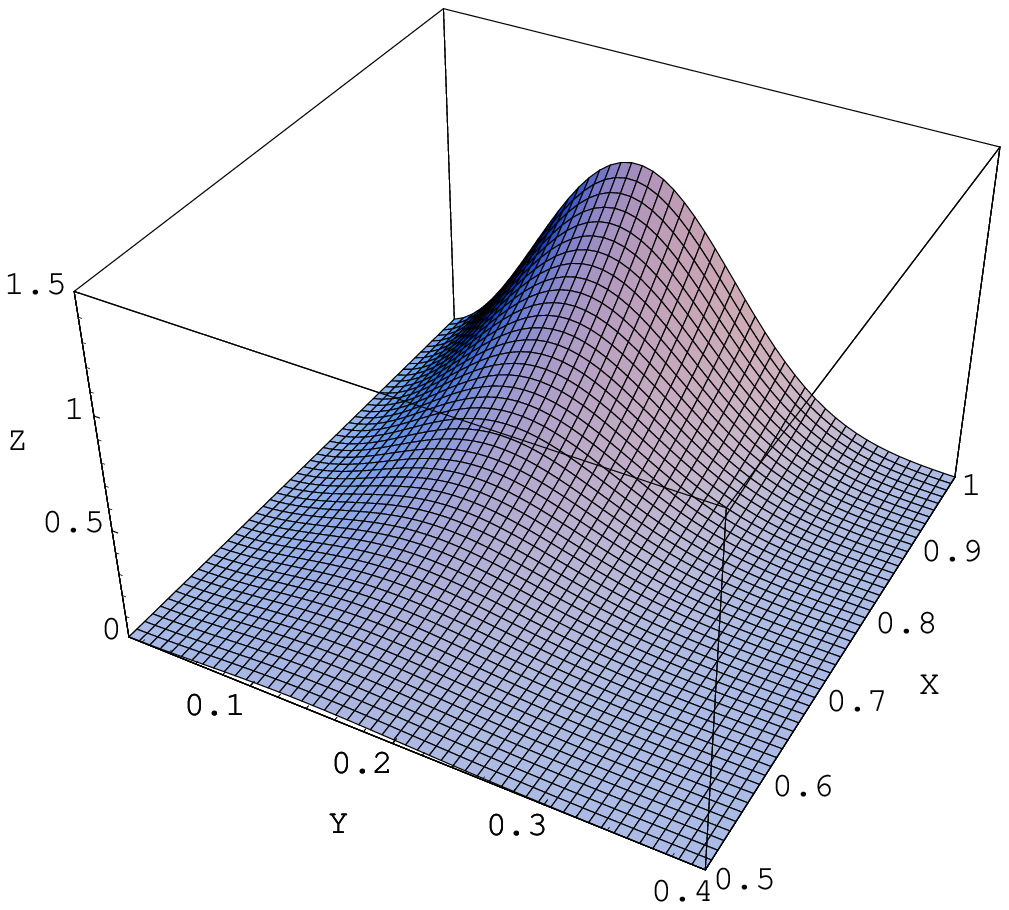,width=6.5cm,height=6cm}& \quad
\epsfig{figure=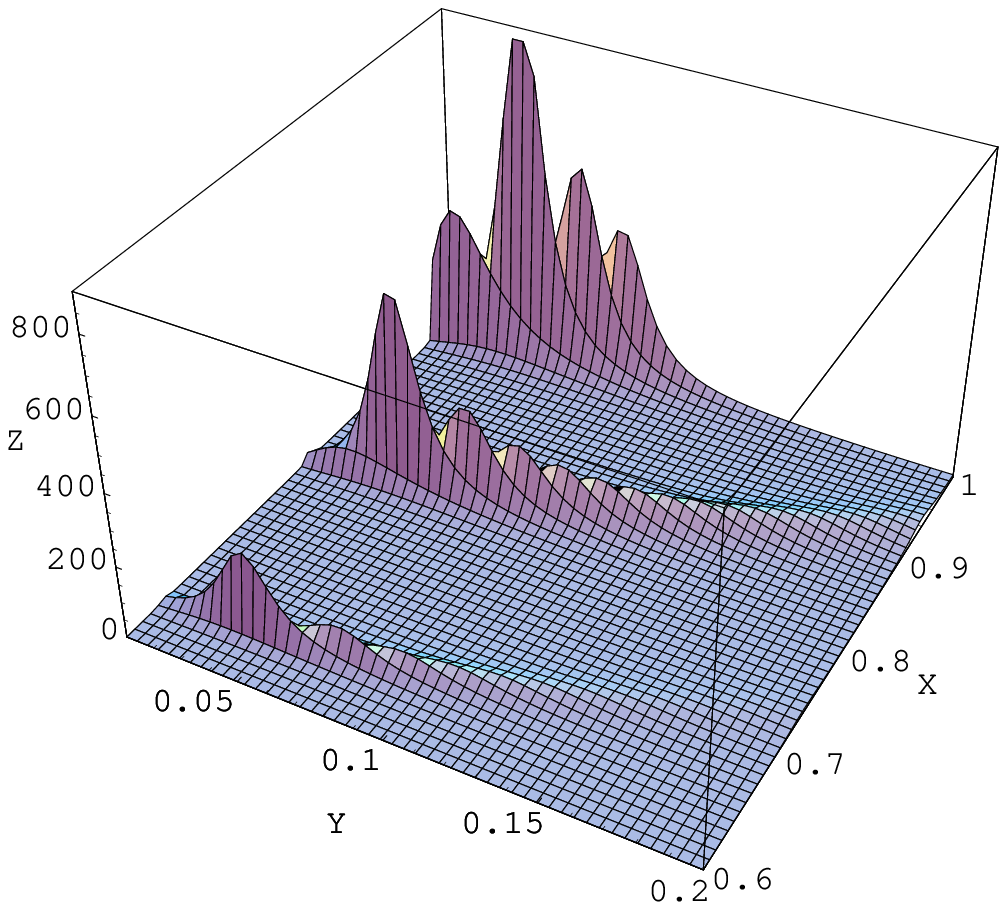,width=6.5cm,height=6cm}
\end{tabular}
\end{center}
\caption{Number of emitted quanta per solid angle, $Z=(\hbar
c/q^2)(2\pi /\omega _0)dN_n/d\Omega $, as a function on $X=\omega'
_0 a'/c $ and $Y=\vartheta $ for $v_0=0.9c$, $n=3$ in vacuum
$\varepsilon _0=\varepsilon _1=1$ (left figure) and in presence of
a cylinder with $\varepsilon _0=3$, $\rho _1=a'$ (right figure).}
\label{fig4mass}
\end{figure}
\begin{figure}[tbph]
\begin{center}
\epsfig{figure=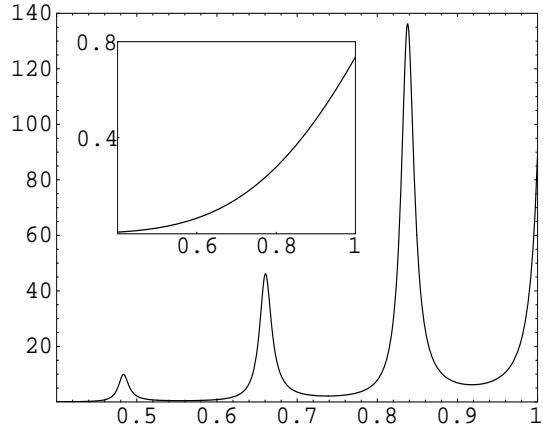,width=9cm,height=6cm}
\end{center}
\caption{The same quantity as in Fig. \ref{fig4mass} as a function
on $\omega' _0 a'/c $ for $\vartheta =0.2 $, $v_0=0.9c$, $n=3$,
$\varepsilon _0=3$, $\rho _1=a'$. The graph in the left-up corner
corresponds to the same quantity for the oscillator radiation in
vacuum ($\varepsilon _0=\varepsilon _1=1$). } \label{fig5}
\end{figure}
In Fig. \ref{fig4mass} we have plotted the number of emitted
quanta (\ref{numerquant1}) as a function on $\omega_0'a'/c$ and
$\vartheta$ for $v_0=0.9c$, $n=3$ in vacuum (left figure) and in
presence of a cylinder with dielectric permittivity
$\varepsilon_0=3$ (right figure). In Fig. \ref{fig5} the number of
radiated quanta is presented as a function on $\omega_0'a'/c$ for
$\vartheta=0.2$. The values for the other parameters are the same
as in Fig. \ref{fig4mass}. As we see from the presented examples
the presence of the dielectric cylinder can lead to the essential
enhancement of the radiation intensity.

\section{Conclusion} \label{sec:Conc}

In the present paper we have considered the influence of a
dielectric cylinder on the electromagnetic radiation from a
longitudinal oscillator with uniform drift along the axis of the
cylinder. By using the Green function, for the general case of the
oscillation law we have derived formulas for the electromagnetic
fields, Eqs. (\ref{electromagneticfield}),
(\ref{electromagneticfield1}), and for the angular-frequency
distribution of the radiation intensity. The latter is a sum of
two terms (see Eq. (\ref{totintOm})). The first one, given by
formula (\ref{Intn0}), is present when the Cherenkov condition for
dielectric permittivity of the exterior medium and drift velocity
is satisfied, $\beta _1>1 $, and describes the radiation
propagating under the Cherenkov angle of the external medium. The
corresponding intensity differs from the radiation intensity of a
charge uniformly moving along the axis of a dielectric cylinder by
presence of the frequency-dependent factor $|g_0(\omega /v_0)|^2$,
where the function $g_n(x)$ is determined by the oscillation law
and is defined by formula (\ref{gnkz}). The second term in the
total radiation intensity (\ref{totintOm}) is given by formula
(\ref{dIndomega}) and describes the radiation, which for a given
propagation direction has a discrete spectrum determined by
formula (\ref{omegantet}). If the Cherenkov conditions for the
maximal and minimal velocities of the charge and dielectric
permittivity of the exterior medium are not satisfied, $|\beta
_{\pm } |<1$, the Cherenkov radiation is absent and for large
values of the harmonic $n$ the radiation intensity is
exponentially suppressed. The corresponding behaviour is
essentially different when $\beta _{+} >1$ or $|\beta _{-} |>1$.
In these cases there are angular regions, defined by relations
(\ref{cases}), there in addition to the radiation from an
oscillating charge one has also the Cherenkov radiation. In these
regions, for large values of the harmonic number the radiation
intensity linearly increases with increasing $n$. However, large
values $n$ correspond to large frequencies for which the
dispersion becomes important. The values of harmonic $n$ with the
maximal radiated intensity are determined by the specific
dispersion law for the dielectric permittivity of the exterior
medium. In presence of a cylinder the radiation intensity at a
given harmonic differs from the corresponding intensity in a
homogeneous medium by the inhomogeneity factor $F_n(\vartheta )$,
defined by formula (\ref{Fntet}). Our calculations for specific
values of the parameters have shown that this factor can
essentially exceed the unity. An example is presented in Fig.
\ref{fig1}. This provides a possibility for the essential
enhancement of the radiation intensity caused by the presence of a
dielectric cylinder. In Sec. \ref{sec:examples} we have specified
the general formula of the radiation intensity for two examples of
the oscillation function. The first one, described in subsection
\ref{subsec:ex1}, corresponds to a charged particle oscillating as
a harmonic oscillator in the laboratory reference frame. The
second example, considered in subsection \ref{subsec:ex2},
corresponds to a charge that executes harmonic oscillations in the
reference frame which moves at a constant velocity. The
corresponding functions $g_n(k_z)$ are determined by formulas
(\ref{gnkzsint}) and (\ref{gnkzprime}). Our numerical calculations
for specific examples have shown that in both these cases the
presence of a dielectric cylinder can essentially increase the
radiation intensity to compared with the radiation in a
homogeneous medium. Note that the results obtained in this paper
are valid also in the case when the oscillator moves inside a hole
along the axis of a dielectric cylinder assuming that the radius
of the hole is much less than the radiated wavelength.

\section*{Acknowledgement}

The authors are grateful to Professor A.~R.~Mkrtchyan for general
encouragement and to Professor L.~Sh.~Grigoryan, S.~R.~Arzumanyan,
H.~F.~Khachatryan for many stimulating discussions. A.~A.~S.
acknowledges the hospitality of the Department of Theoretical
Electrotechnics of the Berlin Technical University and Professor
Heino Henke for helpful discussions and useful comments. The work
has been supported by Grant No.~1361 from Ministry of Education
and Science of the Republic of Armenia.


\bigskip

\begin{thebibliography}{99}

\bibitem{Luch90} P. Luchini and H. Motz, {\it Undulators and
Free-electron Lasers} (Clarendon, 1990).

\bibitem{Brau90} Ch. A. Brau, {\it Free-electron Lasers} (Academic Press,
Boston, 1990).

\bibitem{Nikitin} M. M. Nikitin and V. Ya. Epp, {\it Undulator
Radiation} (Energoatomizdat, Moscow, 1988, in Russian).

\bibitem{Epp} V. Ya. Epp, in {\it Synchrotron
Radiation Theory and Its Developments} edited by V. A. Bordovitsyn
(World Scientific, Singapore, 1999).

\bibitem{Mkrt89} A. R. Mkrtchyan, L. Sh. Grigoryan, A. N. Didenko,
A. A. Saharian, and A. G. Mkrtchyan, Izv. Akad. Nauk Arm. SSR Fiz.
{\bf 24}, 62 (1989)[Sov. J. Contemp. Phys. {\bf 24}, 10 (1989)];
A. R. Mkrtchyan, L. Sh. Grigoryan, A. N. Didenko, and A. A.
Saharian, Sov. Phys. JTP {\bf 61}, 21 (1991); A. R. Mkrtchyan, L.
Sh. Grigoryan, A. A. Saharian, and A. N. Didenko, Acustica {\bf
75}, 1984 (1991).

\bibitem{Saha01} A. A. Saharian, A. R. Mkrtchyan, L. A. Gevorgian,
L. Sh. Grigoryan, and B. V. Khachatryan, Nucl. Instrum. Methods B
{\bf 173}, 211 (2001).

\bibitem{Grigoryan1995} L. Sh. Grigoryan, A. S. Kotanjyan, and A.
A. Saharian, Izv. Akad. Nauk Arm. SSR Fiz. {\bf 30}, 239 (1995)
[Sov. J. Contemp. Phys. {\bf 30}, 1 (1995)].

\bibitem{Grig95b} S. R. Arzumanian, L. Sh. Grigoryan, Kh. V. Kotanjyan,
and A. A. Saharian, Izv. Akad. Nauk Arm. SSR Fiz. {\bf 30}, 106
(1995) [Sov. J. Contemp. Phys. {\bf 30}, 12 (1995)].

\bibitem{Grigoryan1998} L. Sh. Grigoryan, H. F. Khachatryan , and S. R.
Arzumanyan, Izv. Akad. Nauk Arm. SSR Fiz. {\bf 33}, 267 (1998)
[Sov. J. Contemp. Phys. {\bf 33}, 1 (1998)], cond-mat/0001322.

\bibitem{Kot2000} A. S. Kotanjyan, H. F. Khachatryan , A . V. Petrosyan, and
A. A. Saharian, Izv. Akad. Nauk Arm. SSR Fiz. {\bf 35}, (2000)
[Sov. J. Contemp. Phys. {\bf 35}, 1 (2000)].

\bibitem{Kot2001} A. S. Kotanjyan and A. A. Saharian, Izv. Akad.
Nauk Arm. SSR Fiz. {\bf 36},  (2001) [Sov. J. Contemp. Phys. {\bf
36}, 7 (2001)].

\bibitem{Kota02} A. S. Kotanjyan and A. A. Saharian, Mod. Phys.
Lett. A {\bf 17}, 1323 (2002).

\bibitem{KotNIMB} A. S. Kotanjyan, Nucl. Instrum. Methods B
{\bf 201}, 3 (2003).

\bibitem{KhachatryanB} B. V. Khachatryan , Izv. Vissh. Uchebn. Zaved,
 Radiofiz. {\bf 6}, 904 (1963).

\bibitem{Kostanyan1971481} F. A. Kostanyan and O. S. Mergelyan, Izv. Akad.
Nauk Arm. SSR Fiz. {\bf 6}, 481 (1971).

\bibitem{Kostanyan1977} F. A. Kostanyan and O. S. Mergelyan,
Izv. Akad. Nauk Arm. SSR Fiz. {\bf 6}, 472 (1971); {\bf 12}, 179
(1977).

\bibitem{Kuka75} A.~B. Kukanov and E.~K. Suleiman, Izv. Vissh. Uchebn. Zaved,
Radiofiz. {\bf 11}, 7 (1975).

\bibitem{Bars87} K.~A. Barsukov, E.~A. Begloyan, E.~M. Laziev, and
H.~V. Ryazantseva, The radiation from a charged particle, making
oscillations along the direction of motion through the waveguide
(Yerevan Physics Institute, Preprint EFI-964(14)-87 ).

\bibitem{Saha03} A. A. Saharian and A. S. Kotanjyan, Izv. Akad. Nauk
Arm. SSR Fiz. {\bf 38}, 288 (2003).

\bibitem{Abramovitz} {\it Handbook of
Mathematical Functions}, edited by M. A. Abramovitz and I. A.
Stegun (Dover, New York, 1972).

\bibitem{Bolotovsky} B. M. Bolotovsky, Usp. Fiz. Nauk {\bf 75}, 295 (1961).

\bibitem{Prudnikov2} A. P. Prudnikov, Yu. A. Brychkov, and O. I.
Marichev, {\it Integrals and Series, Vol. 2: Special Functions}
(Harwood Academic, Newark, NJ, 1986).

\end{thebibliography}
\end{document}